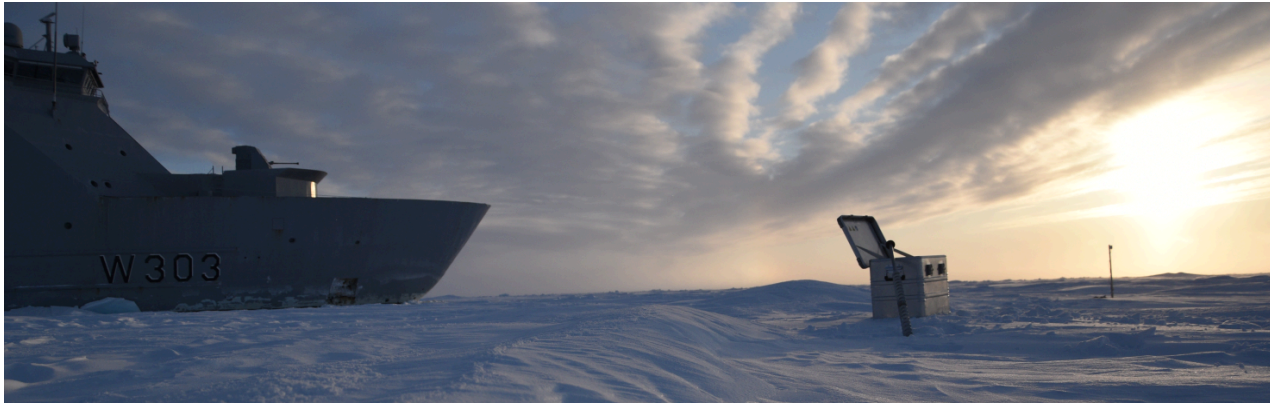

# Cruise Report - KV Svalbard 4.-21.April 2024

## Svalbard Marginal Ice Zone 2024 Campaign

M. Müller[1,2,*] J. Rabault[1], and C. Palerme[1]

(1) Norwegian Meteorological Institute
(2) Department of Geosciences, University of Oslo
(*) corresponding author: maltem@met.no

# 1. Overview

The coupling of weather, sea-ice, ocean, and wave forecasting systems has been a long-standing research focus to improve Arctic forecasting system and their realism and is also a priority of international initiatives such as the WMO research project PCAPS. The goal of the **Svalbard Marginal Ice Zone 2024 Campaign** was to observe and better understand the complex interplay between atmosphere, waves, and sea-ice in the winter Marginal Ice Zone (MIZ) in order to advance predictive skill of coupled Arctic forecasting systems. The main objective has been to set up a network of observations with a spatial distribution that allows for a representative comparison between in situ observations and gridded model data. The observed variables include air and surface temperature, sea-ice drift, and wave energy spectra. With the support of the Norwegian Coast Guard, we participated in the research cruise with KV Svalbard from 4. April - 21.April 2024. In total 34 buoys were deployed in the Marginal Ice Zone north of the Svalbard Archipelago. The first part of the report describes the instruments and their calibration (Section 2), and the second part briefly describes the weather, sea ice, and wave conditions during the campaign. The quality-controlled and calibrated data will be made publicly available as part of a scientific data paper. In addition, the data will be made available on special request to the authors before the final publication.

## 2. Instruments/Measurements

### a. OMB + temperature sensors

The OpenMetBuoy (OMB, *Rabault et al. 2021*) is used as the basis for the instruments deployed. The OMB is an open source, low cost (around 650 USD construction cost all included, and 110 USD per month of activity for iridium cost) buoy with global communication capability through the iridium short burst data global network. The original OMB performs measurements of drift (using GPS, at a sample interval of 30 minutes) and 1-dimensional wave spectrum (using motion sensors, embedded Kalman filtering, and in-situ signal processing; the raw data are sampled at 800Hz for 20 minutes, and a new spectrum is acquired every 2 hours by default). More details are available, e.g., in *Rabault et. al. 2022*, or in previous datasets using the OMB, e.g. *Rabault et. al. 2023.* The OMB GPS and wave measurements are well-validated in the literature. The OMB is adapted to performing measurements in the polar regions and can function for up to 4.5 months in polar temperatures using just 2 non-rechargeable lithium D-cells.

For the *Svalbard Marginal Ice Zone 2024 Campaign*, the OMB has been extended for the first time by four temperature sensors (Fig. 1a). The sensor at one meter height has been mounted on a pole protected by a small 3D-printed open-source radiation shield[1] (*Figure 1b*). Using a low-cost compact radiation shield is necessary to keep costs down when producing a large number of OMBs with temperature measurements. The other sensors are covered with copper tape to allow for good thermal conductivity as well as minimal impact by solar radiation. The two upper sensors were attached to the pole by using a dissolvable string, in order to allow the OMB to release the pole and convert to an ocean buoy when it falls into water and is no longer located on the ice floe.

---

[1] https://www.stlfinder.com/model/temperature-sensor-mount-radiation-shield-volnN5wy/3144546/

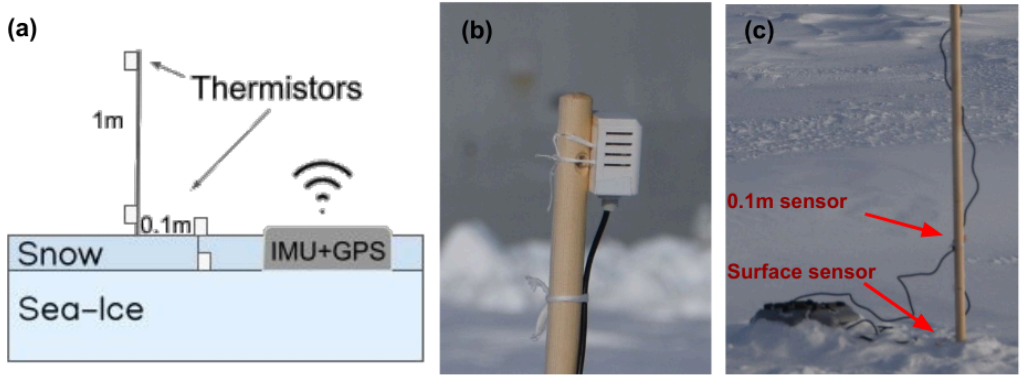

**Figure 1: (a)** OceanMetBuoy (grey box) and four temperature sensors, (1) at the snow-ice interface, (2) at the snow surface, (3) at 0.1-meter height, and (4) at 1-meter height. **(b)** Radiation shield and temperature sensor mounted on a 1-meter pole. **(c)** The buoy set-up on an ice floe.

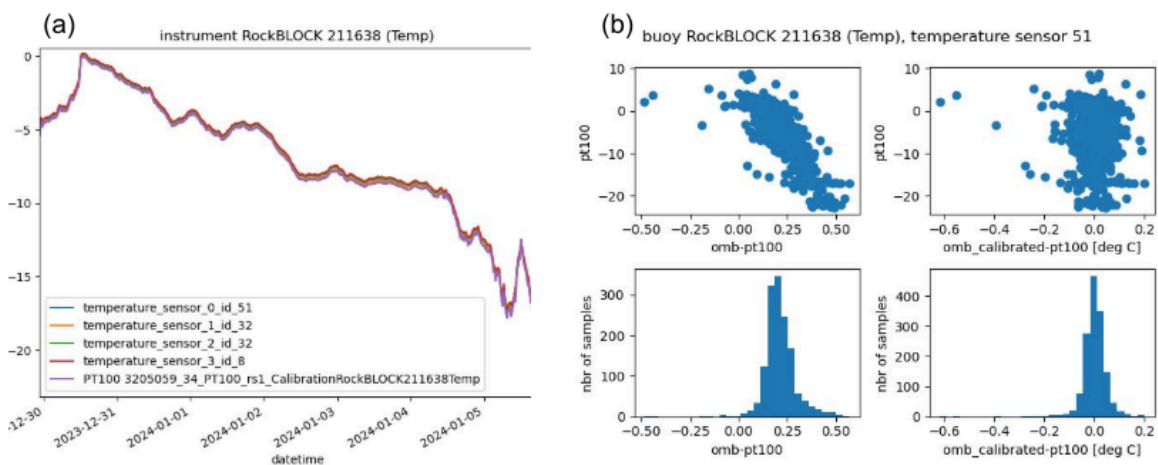

**Figure 2: (a)** Illustration of the timeseries obtained from the PT100 calibrated reference sensor and 4 "raw reading" DS18B20 temperature sensors as obtained from the provider before additional calibration is performed. **(b)** illustration of the custom calibration process. The 2 top figures show the scatter plot of the PT100 vs the temperature sensor 51 before (left) and after (right) linear calibration, and the 2 bottom figures show the corresponding error distributions before (left) and after (right) calibration against the PT100 reference sensor.

The four temperature sensors are DS18B20 digital sensors with 1-wire bus interface. The temperature sensors are individually calibrated against a reference PT-100 thermistor, by putting them all in a professional grade (Stevenson) screen and gathering data for a period of time in the test field of the Norwegian Meteorological Institute in Oslo (ideally, data calibration takes place for several days during the middle of the winter, when cold temperatures are obtained). Taking the calibrated PT100 measurements as a ground truth, and considering that the temperature inside the radiation shield is uniform, and we perform a linear calibration to minimize the mismatch between each DS18B20 temperature sensor and the reference calibrated PT100 probe. We present an example of timeseries of temperature signal used for calibration, alongside scatter plots of the mismatch between the PT100 sensor and one individual DS18B20 temperature sensor before and after calibration, in Figure 2. As visible there, while DS18B20 taken out-of-the-box can have significant offset biases, a DS18B20 sensor with proper linear calibration can have an absolute accuracy of typically slightly below +-0.1 degree C and no offset bias. This is consistent with previously reported results, e.g. The Cave Pearl Program[2].

In addition to the calibration of the sensor per se, several experimental setups have been conducted in the Oslo test field of the Norwegian Meteorological Institute to understand the sensitivities to the radiation shield used for the 1-meter sensor and the thermal conductivity of the surface sensors:

(1) For testing of the radiation shield, we used a reference PT-100 sensor in a standard professional-grade radiation shield and compared the temperature values obtained by the PT100 in the radiation shield, to one calibrated DS18B20 in the small radiation shield. As visible in Fig. 3, the small 3D printed radiation shield has some degree of solar radiation sensitivity compared with the professional grade radiation shield. This is confirmed by looking at Fig. 4, which indicates that some bias due to radiation can take place in certain conditions. Note, however, that these events of strong deviations are rather rare (see distribution from Fig. 3b). The conditions that lead to such deviations will be further analyzed and flagged in the observation data of the campaign.

(2) To test the reliability of the surface temperature sensor we also used the Infrared sensor setup (see Section 2b) and compared it against thermistors on the surface, (1) original white-painted sensor, (2) taped with copper tape, (3) covered with copper wire, (4) aluminum tape (Fig. 5). The results show that the copper tape seems to have the best characteristics in terms of thermal coupling to the surface and is the least impacted by solar radiation. Hence, all surface sensors for the campaign were taped with copper tape.

An additional validation on the sea ice has been conducted with one of the buoys (KVS-12) versus infrared sensor observations, while the ship was attached to the sea-ice floe for about 14 hours (Fig. 6). Generally we find that, while the variability is captured, the surface measurements still need to be taken with care and are not directly comparable to IR measurements since the lowest temperatures are not observed and positive bias occurs.

---

[2] https://thecavepearlproject.org/2016/02/12/triage-step-for-cheap-ds18b20-temperature-sensors/

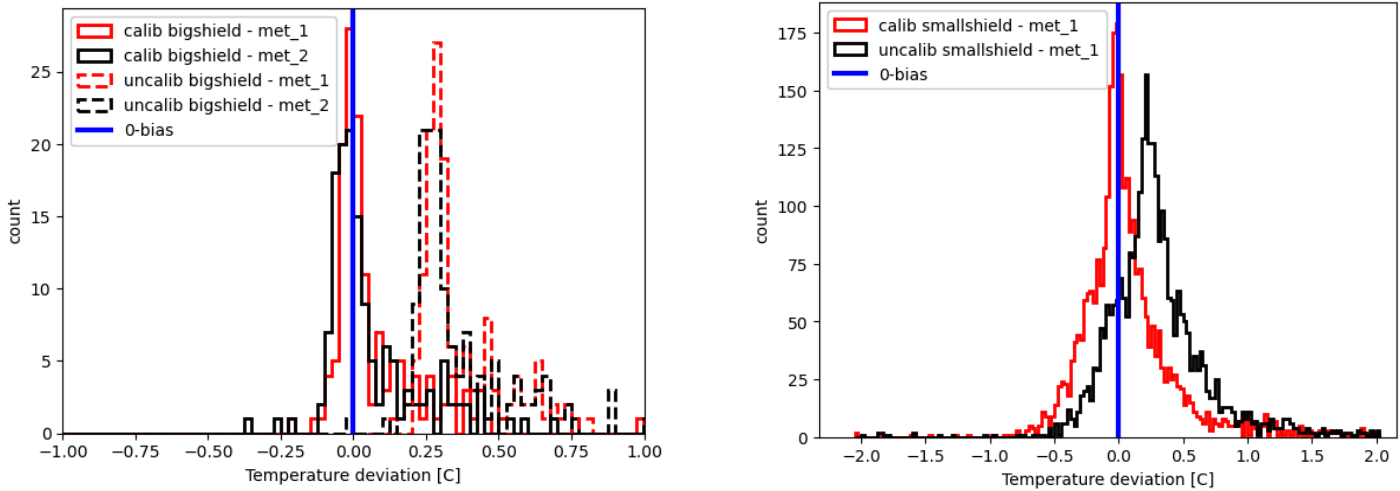

**Figure 3:** Left: illustration of the mismatch between one DS18B20 both calibrated ("calib") and uncalibrated ("uncalib") and 2 different reference PT100 (met_1 and met_2), when all sensors are in the same professional grade radiation shield ("bigshield"), over the period 2024-02-09 to 2024-02-12. The error between the calibrated DS18B20 and the PT100s when all are in the same professional grade shield is then typically within slightly above +-0.1C, consistent with the calibration described above. Right: illustration of the mismatch between one DS18B20 in a small 3D-printed shield used for the buoys ("smallshield"), and a reference PT100 in the professional grade radiation shield (met_1), over the period 14 January to 13 March 2024. The DS18B20 calibration, as expected, can remove systematic temperature biases. However, the small 3D printed radiation shield has different shielding and temperature inertia properties compared with the professional grade shield, and the mismatch tails extend to typically +-0.5C, with an asymmetry of the distribution, indicating some level of solar radiation effect.

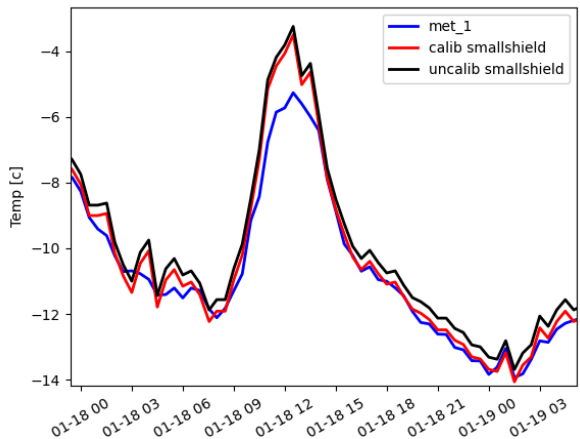

**Figure 4:** Zoomed-in illustration of a bias event due to solar radiation between the PT100 reference temperature sensor in the professional grade shield (met_1), compared to a DS18B20 sensor in the small shield used on the buoys (uncalib smallshield for the uncalibrated raw reading, and calib smallshield for the linearly calibrated reading).

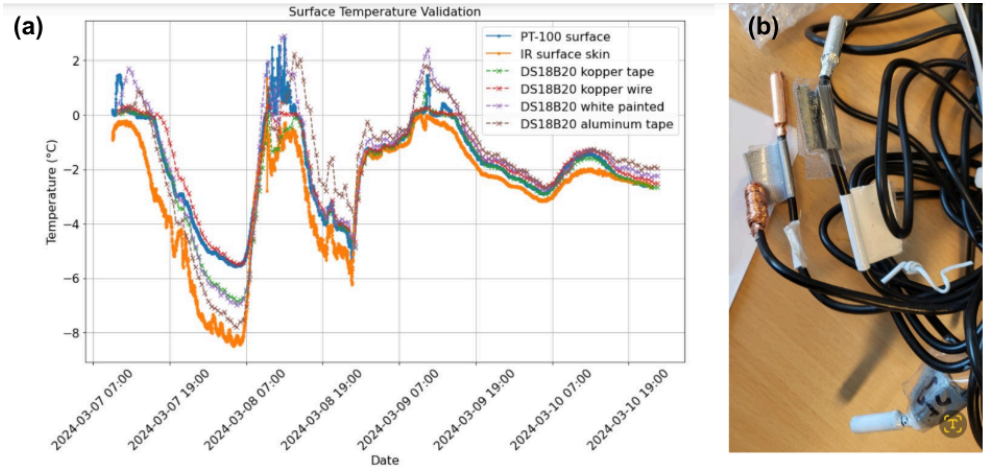

**Figure 5:** Four different versions of the surface temperature sensors have been validated against the PT-100 (on the snow surface) and the Infrared Sensor. In **(a)** the observations from 7 to 10 March 2024 at the Blindern test station, in **(b)** the four different types (1) copper tape, (2) copper wire, (3) white painted, and (4) aluminum tape.

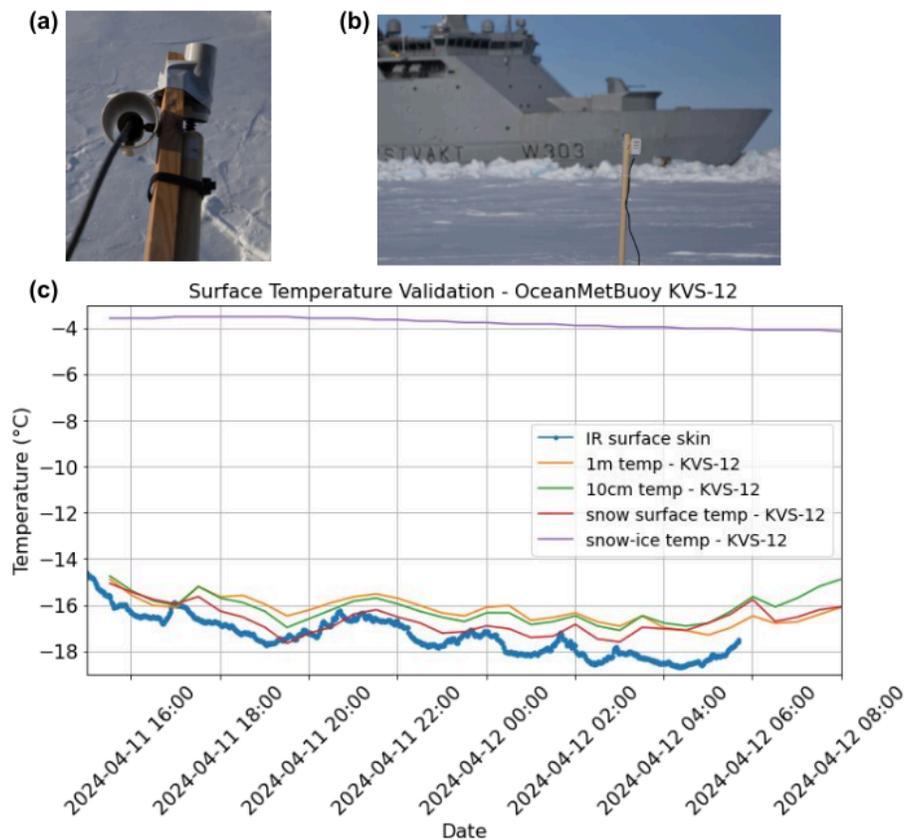

**Figure 6:** In one field experiment, the vessel KV Svalbard was moored to an ice floe where an OpenMetBuoy KVS-12 had been installed, about 200 meters inside the floe, and data have been gathered from 11-April 2024 16Z to 12-April 2024 06Z. **(a)** shows one infrared sensor pointing down to the snow surface and one up to the sky. **(b)** The KVS-12 station. **(c)** The observed temperatures were from the four sensors on KVS-12 and the infrared sensor mounted on the ship.

## b. IR sensor

We have used a mobile infrared sensor setup which has been installed on the starboard side of KV Svalbard at about 5 meters height. The IR sensors are oriented to the surface (Cambell Science IR120) and to the sky (Apogee SI-411). In addition, a reference PT100 thermistor is connected to the logger. The sea-ice surface temperature is calculated by using an atmospheric correction.

## c. Ice thickness, snow depth and density measurements

Sea-ice thickness, snow depth, and density observations were collected at a location close-by (ca. 5 meter) to the OpenMETBuoy station. Two snow depth measurements were performed for each station, one at the buoy location and another one about 5-10 meters from the buoy where we also collected snow cores for assessing the snow density. Vertical sections of the snowpack were performed to measure the snow thickness, and we collected snow cores with a plastic tube with a diameter of 85 mm. Several snow cores were collected at each location to assess the snow density, but the number of snow cores collected per station varies depending on snow depth and time constraints.

## d. OVL Portal

The OceanVirtualLaboratory (OVL, developed by OceanDatalab, Locmaria-Plouzané, France) has been used to monitor the data and to plan the buoy deployment sites during the campaign (Fig. 7). The uncalibrated temperature data and the wave spectra can be assessed via the permalink https://odl.bzh/iCywbjPU and can be visualized in combination with various sources of satellite data information. Data of this campaign can be found under *Products/In Situ/ OpenMetBuoys Svalbard MIZ*.

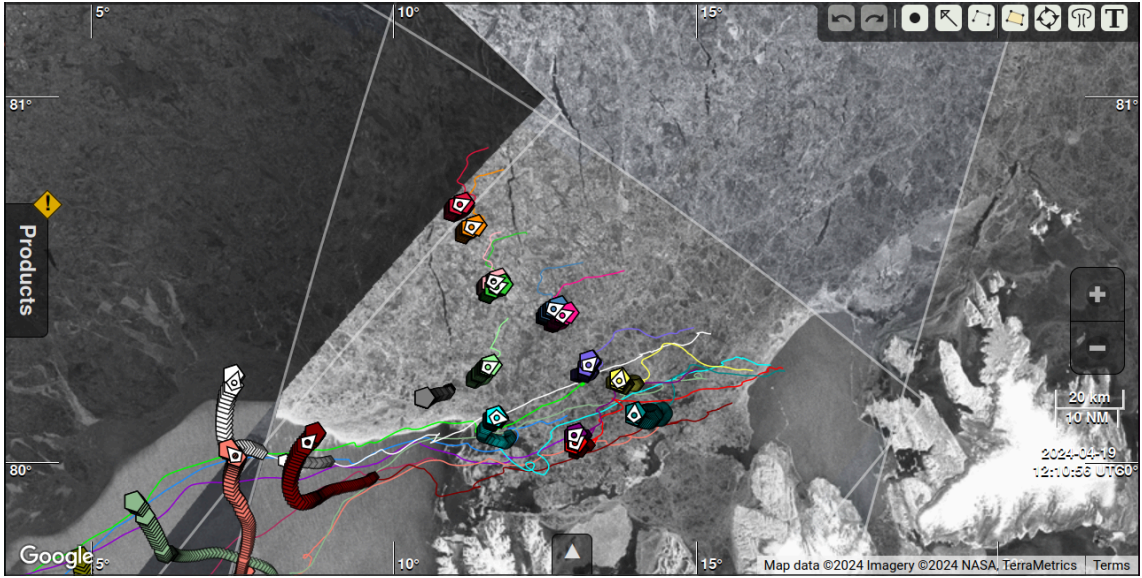

**Figure 7**: Illustration of the OVL web-based visualization of the buoys trajectories on top of SAR images of the area. To interactively browse the data, see https://odl.bzh/iCywbjPU .

# 3. Campaign, results, and data location

## a. Overview

In total, 34 OceanMetBuoys have been successfully deployed during the cruise on the sea-ice North of Svalbard. The original route plan for the campaign has been to the southeast of Svalbard (East of Edgeøya), however, due to the large sea-ice coverage around Svalbard in April 2024 (Fig. 8) the route plan needed to be revised and a route toward the North and North-East of the Svalbard Archipelago was chosen. Due to a persistent Cold Air Outbreak with predominantly northerly and easterly winds the sea ice was very compact in that area, which made it too challenging for KV Svalbard to reach its destination in the northeast of Svalbard (north of Nordaustland). Thus the furthest point to the North East has been in Hinlopenrennet , north of Hinlopen Strait. From that point, the ship returned, with a northward loop out of the Marginal Ice Zone on the 13/14th of April 2024. On that last part of the route, the final 16 buoys have been deployed.

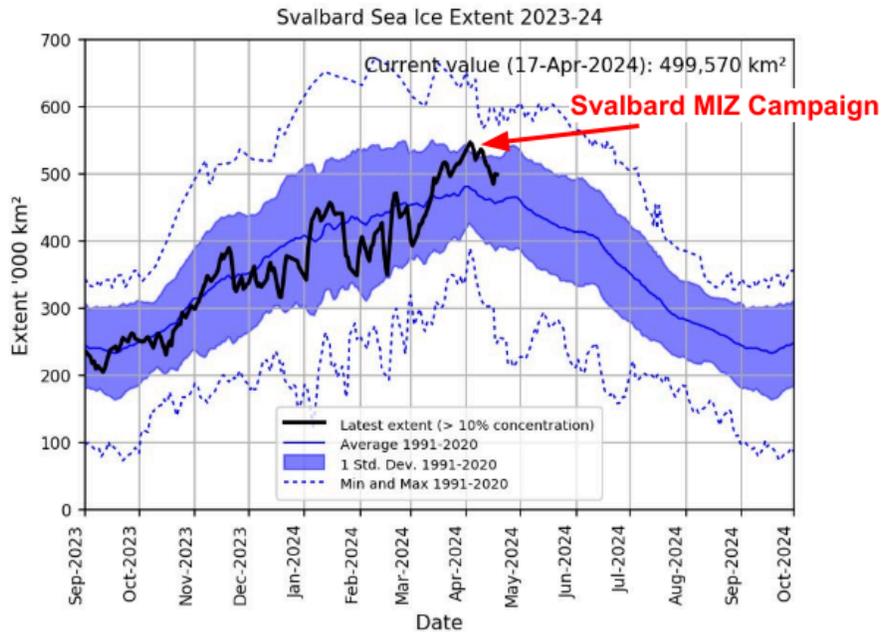

**Figure 8:** Climatology of sea-ice extent around Svalbard and values of the 6 months preceding the cruise. Original figure from cryo.met.no.

## b. Weather, wave, and sea-ice conditions

During the main period of the campaign, from 4th April to the 21st of April, two different weather regimes were dominant. First, from the start of the first deployment, 6th April to 19th April, strong northerly-to-easterly winds with cold temperatures (around -20°C) were dominant. The second period took place after all buoys were deployed in the period from 19th April to 24th April, when the weather changed to southerly winds, in combination with about 20 degree Kelvin higher air temperatures as well as waves of up to 5 meter significant wave height from the Atlantic.

The cold air outbreak with northerly winds before the campaign led to a compact sea-ice cover north of Svalbard, and generally to a positive anomaly in terms of sea-ice coverage around the Svalbard Archipelago (Fig. 9a). In the period from 7th April onward. the easterly winds and the resulting southerly wind through Hinlopen Strait, led to a slowly evolving break-up of the sea-ice and the forming of polynyas north of Spitsbergen and north of Nordaustland (Fig. 9b). The shift in the weather towards southerly warm air and larger waves, finally resulted in a very defined sea-ice edge and northward shift of the sea-ice (Fig. 9c). From April 23 onward the north side of Svalbard was detached from the Marginal Ice Zone, with only a zone of fast ice left (Fig. 9d).

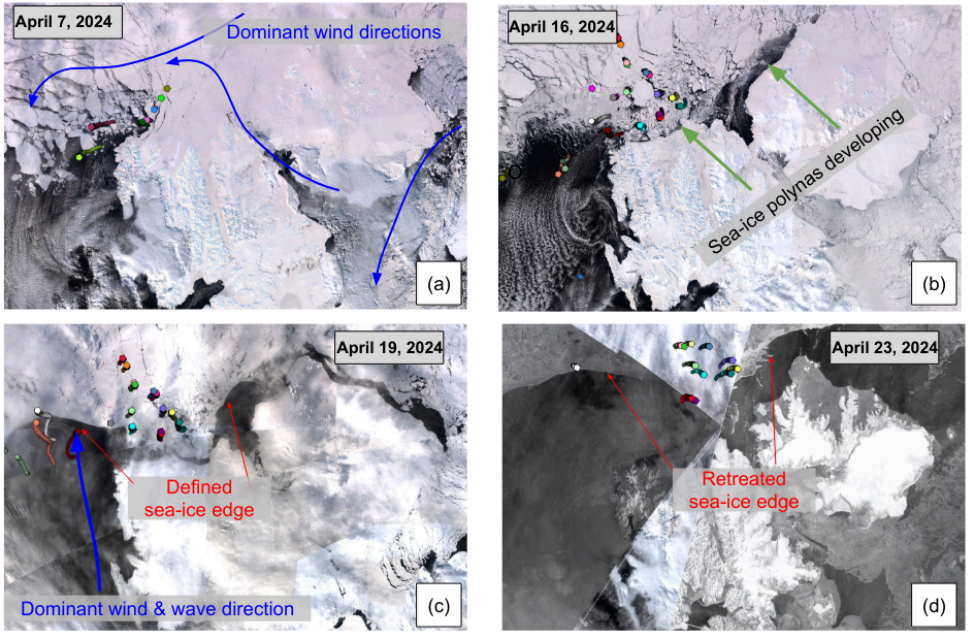

**Figure 9:** Development of the sea-ice conditions through wind and wave forcing from April 7 to 23, 2024.

The drift trajectory and the 1m-temperature observations of the buoy KVS-13 are shown in Figure 10 as an example. In addition, the AROME Arctic weather prediction system's forecast (0-12h lead time) is shown for 0m and 2m-temperature. Many similar features are visible and will be further analyzed in the future.

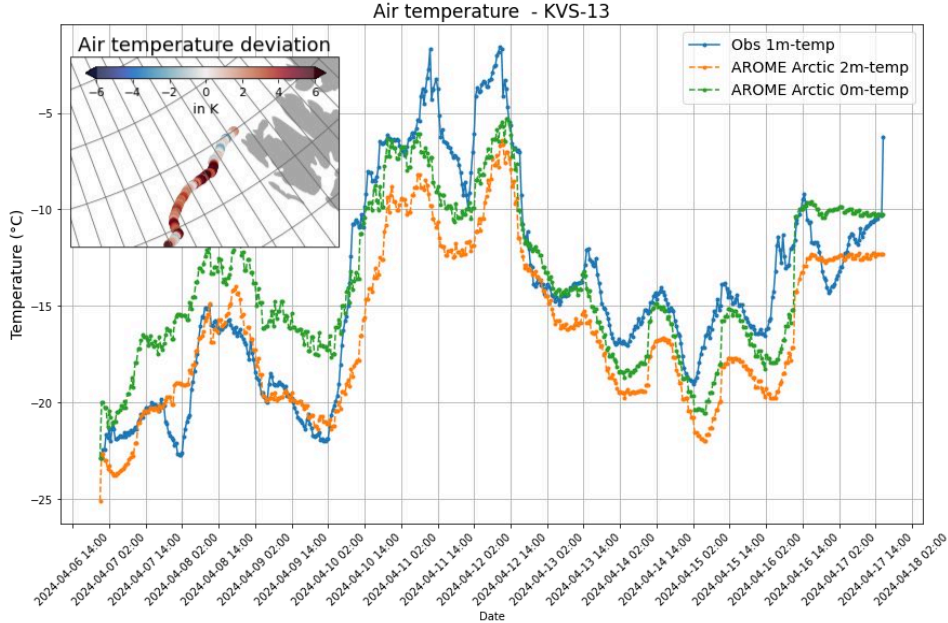

**Figure 10**: 1m-temperature measurements of buoy KVS-13 versus co-located weather forecasts of AROME Arctic (*Müller et al. 2017*) for 0 to 12 hour forecast leadtime.

### c. Ice thickness, snow depth and density distribution.

The sea-ice thickness, snow depth, and density measurements (Table 1) from all stations are shown in FIg. 11.

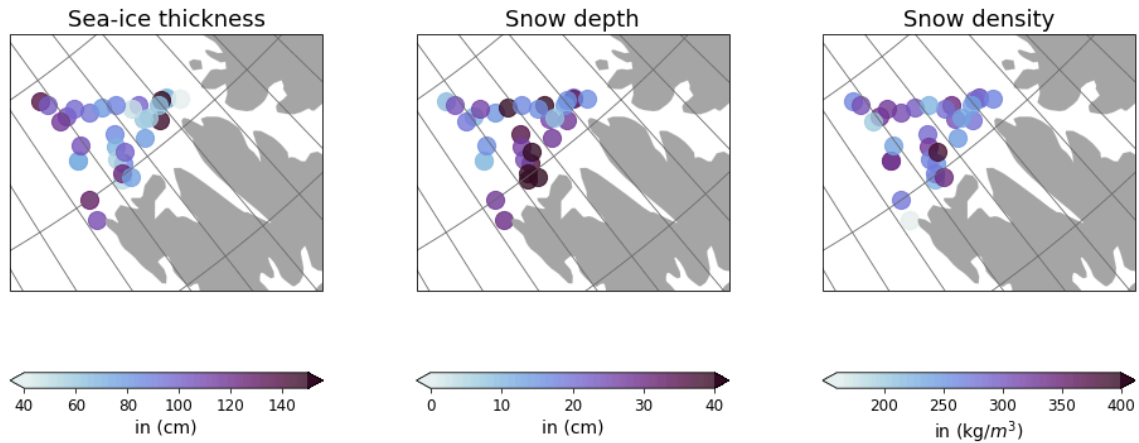

**Figure 11:** The sea-ice thickness, snow depth 2 (see Table 1), and the snow density distribution observed during the deployment of the buoys from April 6 to April 14, 2024.

### d. Data availability

Data will be made available on Github [3] as a netCDF format file following the CF conventions, as provided by the TraJan[4] project, once they are quality controlled. In the meantime, data can be requested directly from the authors.

---

[3] https://github.com/jerabaul29/2024_04_data_release_KV_Svalbard_MIZ_experiment
[4] https://github.com/OpenDrift/trajan

# Appendix

| ID | Date | Time | Snow Depth 1 (in cm) | Snow Depth 2 (in cm) | Ice thickness (in cm) | Snow Density (in kg/m3) |
|---|---|---|---|---|---|---|
| KVS-08 | 2024-04-06 | 11:05:00 AM | 30 | 30 | 110 | 160.17 |
| KVS-13 | 2024-04-06 | 07:42:00 PM | 5 | 28 | 130 | 274.83 |
| KVS-14 | 2024-04-07 | 09:25:00 AM | 48 | 50 | 50 | 221.34 |
| KVS-21 | 2024-04-07 | 10:48:00 AM | 42 | 45 | 125 | 270.21 |
| KVS-18 | 2024-04-07 | 05:00:00 PM | 35 | 25 | 55 | 266.69 |
| KVS-16 | 2024-04-07 | 07:30:00 PM | 18 | 27 | 70 | 326.78 |
| KVS-04 | 2024-04-07 | 10:00:00 PM | 38 | 36 | 87 | 284.08 |
| KVS-22 | 2024-04-08 | 10:15:00 AM | 20 | 48 | 100 | 386.96 |
| KVS-05 | 2024-04-08 | 12:40:00 PM | 25 | 27 | 67 | 263.69 |
| KVS-17 | 2024-04-08 | 05:40:00 PM | 40 | 37 | 87 | 248.78 |
| KVS-31 | 2024-04-09 | 10:20:00 AM | 22 | 42 | 77 | 341.96 |
| KVS-26 | 2024-04-09 | 01:50:00 PM | 13 | 28 | 78 | 243.78 |
| KVS-32 | 2024-04-09 | 03:45:00 PM | 7 | 8 | 57 | 236.58 |
| KVS-09 | 2024-04-09 | 06:00:00 PM | 40 | 48 | 99 | 344.19 |
| KVS-02 | 2024-04-10 | 08:30:00 AM | 40 | 28 | 147 | 278.82 |
| KVS-19 | 2024-04-10 | 10:40:00 AM | 12 | 16 | 160 | 288.35 |
| KVS-03 | 2024-04-11 | 08:30:00 AM | 26 | 19 | 65 | 324.01 |
| KVS-07 | 2024-04-11 | 12:05:00 PM | 27 | 16 | 39 | 263.61 |
| KVS-06 | 2024-04-13 | 09:10:00 AM | 15 | 34 | 65 | 271.86 |
| KVS-27 | 2024-04-13 | 08:45:00 PM | 8 | 15 | 63 | 239.08 |
| KVS-24 | 2024-04-13 | 11:00:00 PM | 25 | 26 | 60 | 305.01 |
| KVS-29 | 2024-04-13 | 11:40:00 PM | 12 | 16 | 47 | 281.23 |
| KVS-34 | 2024-04-14 | 01:15:00 AM | 13 | 17 | 88 | 222.88 |
| KVS-28 | 2024-04-14 | 02:25:00 AM | 35 | 45 | 80 | 297.63 |
| KVS-20 | 2024-04-14 | 03:20:00 AM | 18 | 15 | 93 | 309.77 |
| KVS-25 | 2024-04-14 | 05:20:00 AM | 12 | 26 | 98 | 327.71 |
| KVS-10 | 2024-04-14 | 07:50:00 AM | 8 | 10 | 106 | 334.83 |
| KVS-15 | 2024-04-14 | 10:30:00 AM | 34 | 24 | 103 | 306.93 |
| KVS-11 | 2024-04-14 | 12:00:00 PM | 6 | 9 | 138 | 262.38 |
| KVS-12 | 2024-04-14 | 06:10:00 PM | 8 | 19 | 117 | 199.41 |
| KVS-30 | 2024-04-14 | 09:35:00 PM | 18 | 16 | 106 | 237.03 |
| KVS-35 | 2024-04-14 | 09:35:00 PM | 18 | 16 | 106 | 237.03 |
| KVS-01 | 2024-04-14 | 11:25:00 PM | 3 | 8 | 71 | 334.10 |
| KVS-23 | 2024-04-14 | 11:25:00 PM | 2 | 8 | 71 | 334.10 |

**Table 1:** Overview of the deployment of sensors. **ID** - name of the buoy, **Date and time** of deployment are given in UTC. **Snow depth 1** - measured where the pole with temperature sensor has been mounted. **Snow depth** 2 - mean observed value at the snow measurement location ca. 5-10 meters distant to the pole, **Ice thickness** - measured at a location around 5 meters distant to the pole and snow measurement location. **Snow density** - calculated from the snow probes at the snow measurement location.

# Acknowledgment

This campaign has been funded by the FOCUS (NFR-301450) and by the WMO PCAPS projects. We would like to acknowledge the great support we had from across the Norwegian Meteorological Institute. In particular, Bikas C. Bhattarai and Olaf Weisser for supporting us with the temperature sensor calibration. Eva Sandved for her endless support with the orders of components. Gaute Hope for integrating the buoys into the wavebug system. Steinar Eastwood for loaning us the infrared sensor. Nick Hughes for operational sea-ice information support during the cruise. Jørn Kristiansen and Audun Christoffersen for supporting our ideas and ambitions. We also thank Marius Jonassen, Frank Nilsen, and Lukas Frank from UNIS for their great help with all our logistics issues. In addition, the support from the OVL team (Fabrice Collar / DrFab, Sylvain Herlédan) and the real-time integration of the buoys into the portal has helped us greatly with the planning during the buoy deployment and monitoring the data. The collaboration and support of the KV Svalbard crew have been central to achieving our research objectives. Their expertise and dedication were crucial in navigating through challenging sea conditions and ensuring the safety and effectiveness of our scientific deployments.